# Resolved atomic lines reveal outflows in two ultraluminous X-ray sources

Authors: Ciro Pinto, Matthew J. Middleton, and Andrew C. Fabian

Institute of Astronomy, Cambridge University, Madingley Road, Cambridge CB3 0HA, UK

**Ultraluminous X-ray sources are extragalactic, off-nucleus, point sources in galaxies with an X-ray luminosity above $3 \times 10^{39}$ erg/s, thought to be powered by accretion onto a compact object. Possible explanations include accretion onto neutron stars with strong magnetic fields[1], stellar-mass black holes ($\leq 20$ $M_{solar}$) at or in excess of the classical Eddington limit[2,3,4] or intermediate-mass black holes ($10^{3-5}$ $M_{solar}$)[5]. The lack of sufficient energy resolution in previous analyses has hitherto prevented an unambiguous identification of any emission or absorption lines in the X-ray band, thereby precluding a detailed analysis of the accretion flow[6,7,8]. Here we report the presence of X-ray emission lines arising from highly ionized iron, oxygen and neon with a cumulative significance >5σ, and blueshifted (~0.2c) absorption lines (5σ) in the high-resolution X-ray spectrum of the ultraluminous X-ray source NGC 1313 X-1. In a similar source, NGC 5408 X-1, we also detect emission lines at rest and blueshifted absorption. The blueshifted absorption lines must occur in a fast outflowing gas, whereas the emission lines originate in slow-moving gas around the source. We conclude that the compact object is surrounded by powerful winds with an outflow velocity of about 0.2c as predicted by models of accreting supermassive black holes and hyper-accreting stellar mass black holes[9,10].**

NGC 1313 X-1, NGC 5408 X-1, and NGC 6946 X-1 are three ultraluminous X-ray sources (ULXs, with X-ray luminosities up to ~ $10^{40}$ erg/s). All three sources have been observed in spectral 'states' where a large proportion of the flux emerges below 2keV[6,11] and in these states show strong spectral residuals (0.6-1.2 keV) to the underlying continuum in CCD spectra[8,12]. Due to their relative proximity (D < 7Mpc) and brightness, data quality is high, making them ideal targets for understanding the spectral residuals through the high energy-resolution reflection grating spectrometer (RGS) onboard ESA's XMM-Newton observatory.

NGC 1313 has the lowest star formation rate of the three galaxies and its ULX-1 is well isolated, several arcminutes from the other X-ray bright sources in the galaxy (ULX-2 and SN 1978K, see Extended Data Fig. 1). Archival *Chandra* observations (with sub-arcsec spatial resolution) show it to be confined within a 6 arcsec radius region (<116 pc, at a distance of 3.95 Mpc, see Extended Data Fig. 2). This is not the case for NGC 5408 X-1 whose spectra are affected by a nearby X-ray bright source (see Extended Data Fig. 3a). NGC 6946 has a much high star formation rate, shorter observations and the ULX shows a weaker X-ray continuum (see Extended Data table 1). NGC 1313 X-1 therefore allows for a comparatively 'clean', high energy-resolution study of the features seen in CCD spectra[12].

XMM-Newton has observed NGC 1313 several times in the last 15 years with three observations centred on X-1 each lasting ~100ks, providing independent, well exposed, high-resolution RGS spectra (see Extended Data table 1). We extract the RGS spectra (see Methods for details) and identify strong, rest-frame emission lines from a mixture of elements at varying degrees of ionization including Ne X (12.1 Å), O VIII (19.0 Å) and O VII (21.6 Å) resonance lines (see Fig. 1 and Extended Data Figs. 4 and 5) with the blue side of Ne X partly absorbed. We also find evidence for Fe XVII (15.0-15.3 Å) resonance and (17.1 Å) forbidden lines. We apply a series of physically consistent models for the lines to both the EPIC and RGS data *simultaneously* (see Methods)[13]. The residuals to the blackbody and power-law continuum model previously seen in bright ULXs[8,12] are now resolved by the inclusion of the RGS data into a complex of emission and absorption lines (see Fig. 1). The emission lines are highly significant (at > 3σ each for Ne X, Fe XVII, and O VIII and > 5σ in total) and can be well modelled with a rest-frame, collisionally-ionized gas, which includes an underlying weak bremsstrahlung continuum at an average temperature of 0.8 keV (~ $10^7$ K, see Fig. 1 and Extended Data table 2). The absorption lines (significant at 5σ in total) can be well modelled with two-phase, low ionization absorbing gas in photoionization equilibrium applied to the continuum. One absorber is consistent with being at rest, while the other requires a high outflow velocity around 0.2$c$. More details on the spectral modelling (and abundance ratios) are reported in the Methods. The inclusion of a third, velocity-broadened absorber[8] significantly improves the fit relative to the continuum model. This model requires moderately relativistic velocities (~ 0.25$c$), a high column density ($N_H$ ~ $1 \times 10^{24}$ cm$^{-2}$) and high ionization parameter (ξ ~ $3 \times 10^4$ erg cm s$^{-1}$). It is described by the blue line in Fig. 1. The firm detection and identification of rest-frame emission and blueshifted absorption lines opens up new powerful means to understand ULXs.

The individual observations of NGC 1313 X-1 show evidence for line variability (see e.g. Ne X in Extended Data Fig. 5). Absorption is detected in the first two observations, while the emission lines are stronger in observation 3 where their flux is twice than that seen in observation 1 (see Methods). Emission lines are weaker in observation 2 and show a decrease in the ionization parameter. We do not detect significant absorption in observation 3. In Fig. 2 we show the ratios of the RGS spectra between the individual exposures, which confirm the variability of both absorption and emission lines. We do not find a significant trend in the strength of the features with the spectral hardness of the source most likely because the spectral states of the three observations are very similar[12].

Spectral fits performed using only the RGS data (i.e. excluding the PN data) from the three 100 ks exposures confirmed the detection of the emission and absorption components (see Methods). In Fig. 3 we show the significance obtained adopting 500 km s$^{-1}$ and 10,000 km s$^{-1}$ line widths; negative values indicate absorption lines. Each emission line is detected *individually* at 3σ confirming the > 5σ detection obtained with the CIE emission model (which treats all of the lines consistently). Ne IX and X blueshifted absorption is also individually detected between 3-5σ in agreement with the photoionization code.

The emission lines show comparable fluxes of $2.5 \times 10^{-6}$ ph s$^{-1}$ cm$^{-2}$ and equivalent widths (EW) of 15-30 mÅ; whilst the absorption lines have EW from 15 mÅ up to 250 mÅ. Narrow line region (NLR) lines of typical warm absorbers that are found around active galactic nuclei (AGN) have similar EW[14], but different dominant species: the O VII and Ne IX triplets and the common Fe XVII unresolved transition array (15-17Å)[15]. Galactic X-ray binaries and microquasars also show comparable EW[16,17] and similar ionic species, e.g. Ne X and O VIII, but intercombination transitions may play an important role[18]. The emission lines in NGC 1313 X-1 differ from those typically seen in AGN and some X-ray binaries, but they are very similar to those produced by the accretion disk of the X-ray binary 4U 1626-67[18], which may suggest an origin from a wind launched by a compact disk. The features detected in NGC 1313 X-1 are more difficult to detect than those seen in X-ray binaries because of the several orders of magnitude difference in distance (from a kiloparsec scale for the binaries in our Galaxy to the megaparsec scale for NGC 1313 X-1).

Whilst the emission lines are seen at their rest-frame energies, the blueshifted absorption lines confirm the presence of an outflow, i.e. photoionized gas within a wind[8,12]. The high ionization parameter and outflow velocity suggests an accretion-disk-wind origin similar to that of Galactic black hole binaries[15], but with far larger velocities and therefore energetics (see Methods for possible interpretations). The fact that we detect both emission and absorption lines - the latter being line-of-sight dependent - demands that NGC 1313 X-1 must be seen at a moderate inclination angle (assuming an equatorial wind)[12] with the difference between the widths and the Doppler shifts of the emission and absorption lines suggesting different spatial locations.

If collisional equilibrium applies, the most obvious explanation for the emission lines is either shock/collisional heating in the outflow with a range of velocities or between the outflow and that of the stellar companion as commonly seen in colliding-wind binaries[19,20]. Such lines should also be present in other ULXs, but absorption features may be harder to detect if there is source confusion or if the spectral 'state' is hard[12].

NGC 5408 X-1 also shows sharp emission features similar to those detected in NGC 1313 X-1 including the Ne X and the O VIII resonance lines and narrow absorption features at 12.2 Å, 15.5 Å and 17-18 Å (see Fig. 4, Extended Data Fig. 6, and Methods for details). The strongest emission lines are individually detected at 5$\sigma$, whilst the absorption features have lower significance (about 4$\sigma$ in total). Most features can be described by an isothermal emission model of gas in collisional ionization equilibrium (T ~ 3 keV) and a relativistically-outflowing (v = 0.22$c$) photoionized gas model (see red line in Fig. 4). The main absorbers in NGC 5408 X-1 and NGC 1313 X-1 have comparable outflow velocities (v ~ 0.2$c$), suggesting that they could have the same origin although the wind may be more structured in NGC 5408 X-1 (see Methods).

NGC 6946 X-1 exhibits the O VIII (19.0 Å) and Ne IX (13.45 Å) emission lines and a feature at ~ 11.0 Å that could be attributed to either higher ionization Fe XXII-XXIII emission lines or to an absorption edge (see Extended Data Fig. 4). Its very low continuum prevents the detection of any absorption lines.

The emission lines appear in all three ultraluminous X-ray sources and are likely associated with collisional shock heating between the circumsystem gas and the outflowing wind which we have now identified in NGC 1313 X-1 and NGC 5408 X-1. This result suggests that the accretion flow in some ultraluminous X-ray sources can be associated with powerful winds which leave their imprint in emission and absorption lines and are able to produce the common residuals in the high-quality CCD-resolution spectra of the most bright, well studied ULXs, e.g. NGC 1313 X-1, Ho IX X-1, Ho II X-1, NGC 55 X-1, NGC 5204 X-1, NGC 5408 X-1, and NGC 6946 X-1[12].

**References**

[1] Bachetti, M; Harrison, F. A.; Walton, D. J. et al. An ultraluminous X-ray source powered by an accreting neutron star, *Nature*, **514**, 7521, p. 202-204 (2014)

[2] Shakura, N. I. & Sunyaev, R. A. Black Holes in Binary Systems: Observational Appearance, *Astron. Astrophys.*, **24**, p. 337-355 (1973)

[3] Poutanen, J.; Lipunova, G.; Fabrika, S. et al. Supercritically accreting stellar mass black holes as ultraluminous X-ray sources, *Mon. Not. R. Astron. Soc.*, **377**, p. 1187-1194 (2007)

[4] King, A. R.; Davies, M. B.; Ward., M. J. et al. Ultraluminous X-Ray Sources in External Galaxies, *Astrophys. J.*, **552**, p. L109-L112 (2001)

[5] Pasham, D. R.; Strohmayer, T. E. & Mushotzky R. F. A 400 solar mass black hole in the Ultraluminous X-ray source M82 X-1 accreting close to its Eddington limit, *Nature*, **513**, p.74-76 (2014)

[6] Stobbart, A.-M.; Roberts, T. P.; Wilms, J. XMM-Newton observations of the brightest ultraluminous X-ray sources, *Mon. Not. R. Astron. Soc.*, **368**, p. 397-413 (2006)

[7] Bachetti, M.; Rana, V.; Walton, D. J. et al. The Ultraluminous X-Ray Sources NGC 1313 X-1 and X-2: A Broadband Study with NuSTAR and XMM-Newton, *Astrophys. J.*, **778**, p. 163-173 (2013)

[8] Middleton, M. J.; Walton, D. J.; Roberts, T. P. & Heil, L. Broad absorption features in wind-dominated ultraluminous X-ray sources?, *Mon. Not. R. Astron. Soc.*, **438**, p. L51-L55 (2014)

[9] King, A. & Muldrew, S. I. Black Hole Winds II: Hyper–Eddington Winds and Feedback, *Mon. Not. R. Astron. Soc.*, **455**, p. 1211-1217 (2015)

[10] King, A. & Pounds, K. Powerful Outflows and Feedback from Active Galactic Nuclei, *Annual Review of Astronomy and Astrophysics*, **53**, p. 115-154 (2015)

[11] Gladstone, J. C.; Roberts, T. P.; Done, C. The ultraluminous state, *Mon. Not. R. Astron. Soc.*, **397**, p. 1836-1851 (2009)

[12] Middleton, M. J.; Walton, D. J.; Fabian, A. C.; et al. Diagnosing the accretion flow in ULXs using soft X-ray atomic features, *Mon. Not. R. Astron. Soc.*, **454**, p. 3134-3142 (2015)

[13] http://www.sron.nl/spex

[14] Kaspi, S.; Brandt, W. N.; George, I. M. et al. The Ionized Gas and Nuclear Environment in NGC 3783. I. Time-averaged 900 Kilosecond Chandra Grating Spectroscopy, *Astrophys. J.*, **574**, p. 643-662 (2002)

[15] Ponti, G.; Fender, R. P.; Begelman, M. C. et al. Ubiquitous equatorial accretion disc winds in black hole soft states, *Mon. Not. R. Astron. Soc.*, **422**, p. 11-15 (2012)

[16] Miller, J. M.; Raymond, J.; Reynolds, C. S. et al. The Accretion Disk Wind in the Black Hole GRO J1655-40, *Astrophys. J.*, **680**, p. 1359-1377 (2008)

[17] Marshall, H. L.; Canizares, C. R. and Schulz N. S. The high-resolution x-ray spectrum of ss 433 using the chandra hetgs, *Astrophys. J.*, **564**, p. 941-952 (2002)

[18] Schulz, N. S.; Chakrabarty, D.; Marshall, H. L. et al. Double-peaked X-Ray Lines from the Oxygen/Neon-rich Accretion Disk in 4U 1626-67, *Astrophys. J.*, **563**, p. 941-949 (2001)

[19] Cooke, B. A., Fabian, A. C. & Pringle, J. E. Upper limits to X-ray emission from colliding stellar winds, *Nature*, **273**, p. 645-646 (1978)

[20] Oskinova, L. M. Evolution of X-ray emission from young massive star clusters, *Mon. Not. R. Astron. Soc.*, **361**, p. 679-694 (2005)

[21] Pakull, M. W. & Mirioni, L. Optical counterparts of ultraluminous x-ray sources, *in the proceedings of the symposium 'New Visions of the X-ray Universe in the XMM-Newton and Chandra Era'*, pp. 8 (2002), http://adsabs.harvard.edu/abs/2002astro.ph..2488P

**Author contributions:** CP wrote the manuscript with comments from all the authors and analysed the XMM-Newton data. Both MM and ACF made significant contributions to the overall science case and manuscript.

**Competing financial interests:** The authors declare no competing financial interests.

**Corresponding author:** All correspondence should be addressed to CP. (cpinto@ast.cam.ac.uk)

**Acknowledgements:** The authors thank Anne Lohfink for her help with the use of Photoionization emission codes. The authors also thank the two referees for their useful comments and suggestions which improved the quality and the clarity of the paper. ACF thanks the ERC Advanced Grant on Feedback 340492. MJM appreciates support via an STFC advanced fellowship. This work is based on observations with XMM-Newton, an ESA science mission with instruments and contributions directly funded by ESA Member States and NASA. This research has also made use of data obtained from NASA's *Chandra* satellite. All codes used are publicly available.

**Figure 1: Simultaneous spectral fits to the stacked XMM-Newton RGS and EPIC/PN spectra of NGC 1313 X-1.**
The main frame shows the RGS stacked spectrum; the PN stacked spectrum is in the small insert. The rest-frame wavelengths of the most relevant transitions and some blueshifted lines are labelled. An isothermal emission model of gas in collisional ionization equilibrium describes most emission lines at rest. The absorption lines are reproduced with multi-phase models for gas in photoionization equilibrium. The red line model consists of rest-frame absorption and emission and a relativistically-outflowing (v = 0.2c) photoionized absorber. The blue line is a model which includes an additional broadened absorber (v = 0.25c). Errorbars are given at the $1\sigma$ level.

**Figure 2: Ratios between the individual RGS spectra of NGC 1313 X-1.**
The RGS spectra were normalized by the spectral continuum and divided by that of observation 1. The absorption features (10.7-11.4Å) change in Observation 3. Rest frame emission features also exhibit variability (see also Extended Data Fig. 5). Errorbars are given at the $1\sigma$ level.

**Figure 3: Significance of the features in the NGC 1313 X-1 RGS stacked spectrum.**
The line significance obtained by Gaussian fitting over the 7-27 Å wavelength range with increments of 0.05 Å and negative values indicating absorption lines. The solid and dashed lines indicate the line significance obtained with 500 km s$^{-1}$ and 10,000 km s$^{-1}$ widths, respectively. Ne X, Fe XVII, and O VIII emission lines are individually detected at $3\sigma$ which combined provide an $8\sigma$ detection. Ne X blueshifted absorption is clearly detected up to $5\sigma$ showing widths larger than the emission lines.

**Figure 4: Best fit to the stacked XMM-Newton RGS spectrum of NGC 5408 X-1.**
Line labels are same as in Fig. 1. An isothermal emission model of gas in collisional ionization equilibrium describes most emission lines at rest. The absorption lines can be reproduced with gas in photoionization equilibrium. The red line is a model consisting of single, relativistically-outflowing (v = 0.22$c$), photoionized absorber. The blue line is a model which includes two absorbers ($v_1$ = 0.10$c$ and $v_2$ = 0.22$c$). The absorption lines have a widths of $500 \pm 300$ km s$^{-1}$, while the emission lines are broader with $\sigma_v = 2000 \pm 500$ km s$^{-1}$. Errorbars are given at the $1\sigma$ level.

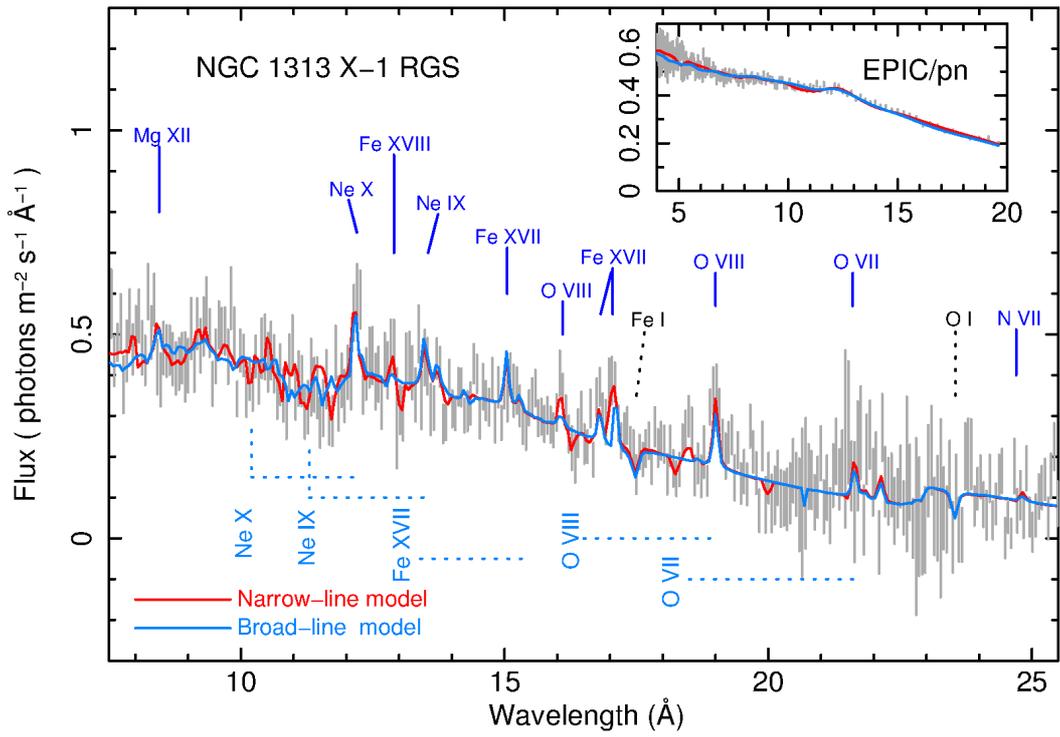

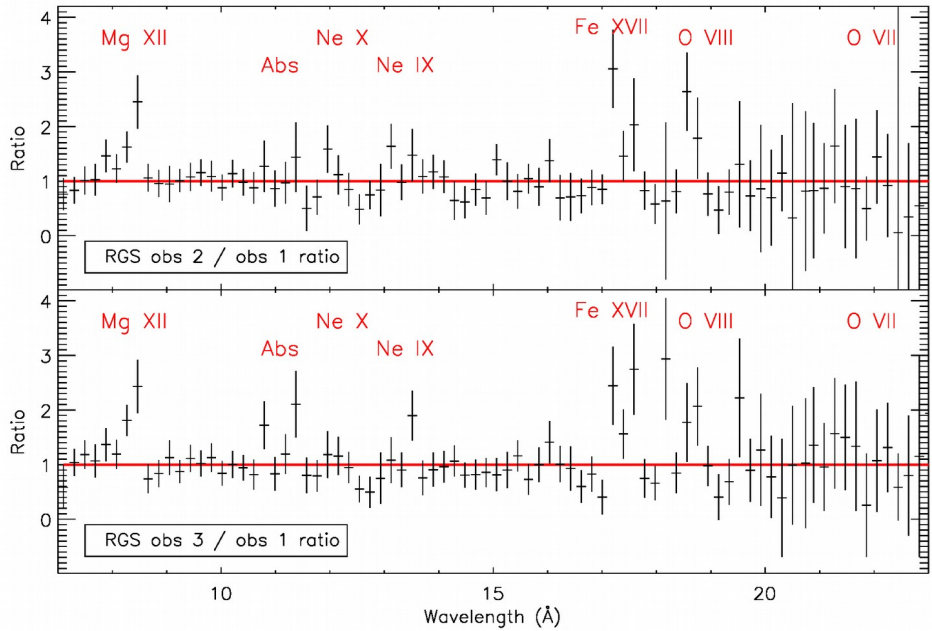

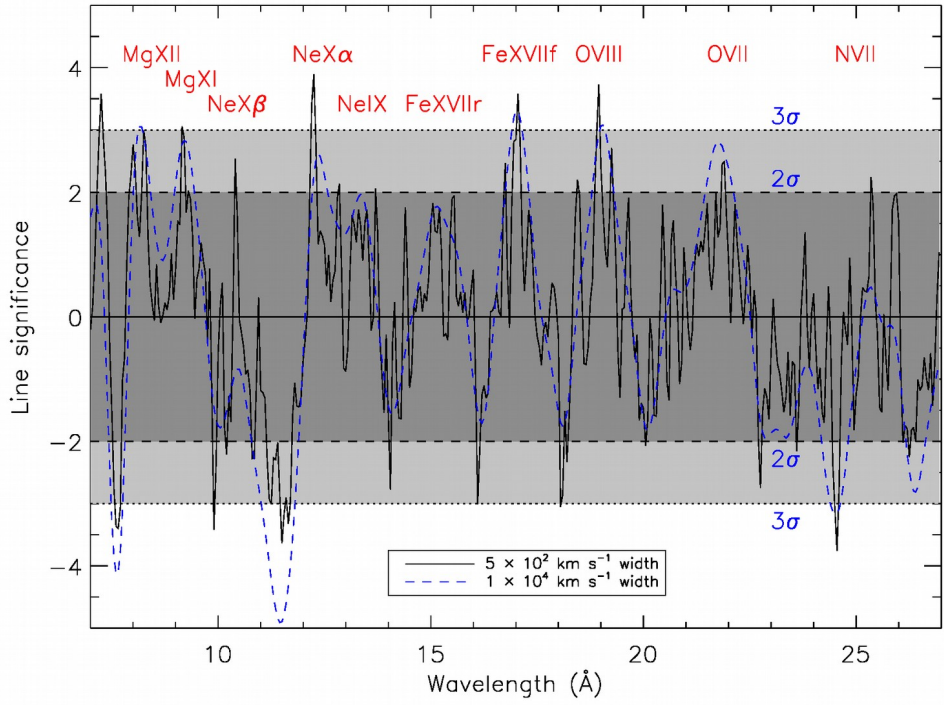

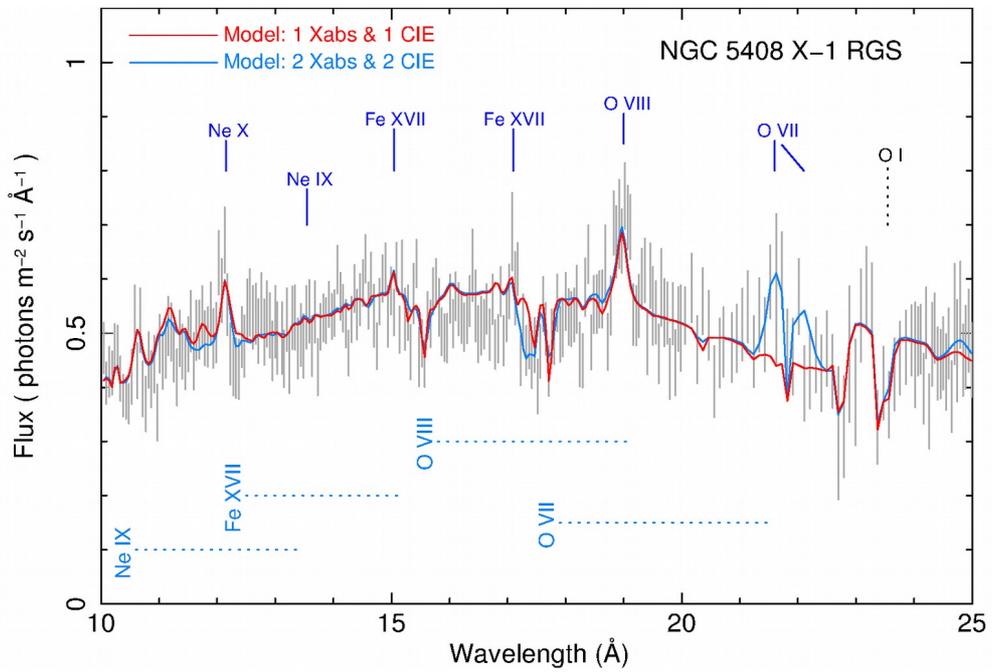

## Methods

### Data reduction

The XMM-Newton satellite is equipped with two types of X-ray detectors: the CCD-type European Photon Imaging Cameras (EPIC)[22,23] and the Reflection Grating Spectrometers (RGS)[24]. The European photon imaging cameras are MOS and PN. The RGS camera consists of two similar detectors, which have high effective area and high spectral resolution between 6 and 38 Å.

All the observations of the sources have been reduced with the XMM-Newton Science Analysis System (SAS) v13.5.0[25]. We correct for contamination from soft-proton flares following the XMM-SAS standard procedures. For each source and exposure, we extracted the first-order RGS spectra in a cross-dispersion region of 1 arcmin width, centred on the emission peak. We have extracted background spectra by selecting photons beyond the 98% of source point-spread-function. The background spectra were comparable to those from blank field observations. We extracted the MOS and PN images in the RGS (0.35-1.8 keV) energy band and stacked them all with the *emosaic* SAS task (see Extended Data Figs. 1 and 3). We also extracted EPIC MOS and PN spectra from within a circular region of 1 arcmin diameter centred on the emission peak. The background spectra were extracted from within a 1 arcmin circle in a nearby region on the same chip, but away from bright sources and the readout direction. As the EPIC/PN spectra contain the majority of the counts and the residuals have been shown to not be instrumental in origin[26], we discard the EPIC/MOS spectra from our analysis. The total clean exposure times are quoted in Extended Data Table 1.

### EPIC + RGS Spectral modelling

We fit the EPIC/PN and RGS spectra (with the SPEX package[13]) *simultaneously* to constrain both the broad-band continuum and describe the atomic features. Importantly, we fit across individual spectra in each observation rather than stacking the data in order to avoid any spurious features resulting from different pointing and background subtraction, which differ between RGS 1 and 2. We bin both the RGS and PN spectra in channels equals to 1/3 of the PSF, and use C-statistics, because it provides the optimal spectral binning and avoids over-sampling.

The phenomenological continuum model we apply to the data is a combination of soft blackbody with temperature $T \sim 2.5 - 3.0 \times 10^6$ K and power-law emission components with photon index $\Gamma \sim 1.9$ extending to high energies, both absorbed by neutral gas with a best-fit hydrogen column density $N_H = 1.8 \pm 0.1 \times 10^{21}$ cm$^{-2}$, which includes any intrinsic and Galactic absorption ($N_H^{Gal} = 4 \times 10^{20}$ cm$^{-2}$)[27] adopting solar abundances[28]. The continuum of the three NGC 1313 X-1 observations shows little evidence for variability above 1 keV while the soft X-ray band is variable. All the model parameters for the RGS and EPIC spectra of the same observation are tied with each other whilst the continuum parameters (normalization and slope of the power law, normalization and temperature of the blackbody) are uncoupled between different observations. We detect a complex of emission and absorption lines (see Fig. 1). Under the first-order assumption that the lines are unchanging between observations we also tie the parameters of the absorption and emission-line models for the different observations in order to increase the statistics.

The emission lines resolved by the RGS can be well modelled with a rest-frame, collisionally-ionized gas (CIE model in SPEX), which is detected up to 8σ (see Fig. 1, red line, and Extended Data table 2). The absorption-like features can be modelled with a two-phase absorbing gas in photoionization equilibrium. This can be described by a combination of two XABS models in SPEX. The $\Delta C_{stat}$ and the equivalent $\Delta \chi^2$ provided by each component, which indicates their improvement to the fit, are reported in Extended Data table 2. The ionization parameters ξ were tied between the different XABS models because a preliminary fit showed that, if they are left free to vary, they agree within error. One component is consistent with being at rest, while the other requires a high outflow velocity ~ 0.2$c$ ($v_2$ = -65000 ± 10500 km s$^{-1}$). The column densities of the two absorbers are $N_{H,1}$ = 1.5 ± 0.3 × 10$^{22}$ cm$^{-2}$ and $N_{H,2}$ = 5 ± 1 × 10$^{21}$ cm$^{-2}$, respectively. At this stage, the velocity broadening was tied between the emission and the absorption components with the best fit giving a 1σ upper-limit of about 500 km s$^{-1}$. When we untie the velocity broadening between these components, we obtain $v_{\sigma,CIE}$ = 1000 ± 500 km s$^{-1}$ and $v_{\sigma,XABS}$ < 20 km s$^{-1}$, respectively. The ionization parameter of the absorbers is rather low (ξ = 200 ± 100 erg cm s$^{-1}$).

It is not possible to determine the absolute metal abundances from the RGS lines, as these require comparison to hydrogen lines which are absent in X-ray spectra. In the fits above, the elemental abundances were therefore tied between the emitting and absorbing gas components at solar metallicity for iron[28] with free ratios for the abundances of oxygen, neon, and magnesium (because Fe XVII and Fe XVIII produce detectable lines: see Fig. 1). On average, the emitting and absorbing plasmas exhibit abundance ratios: O/Fe = 1.0 ± 0.2, Ne/Fe = 1.8 ± 0.4, and Mg/Fe = 2.1 ± 0.5. This would suggest a small over-abundance of α elements (with respect to iron), which indicates a small amount of supernovae core-collapse (SN cc) enrichment with respect to the solar ratio of SN Ia to SN[28,29]. Most ULXs - including the three studied in this work - are found in spiral / interacting / star-forming galaxies where a recent contribution from SN cc could be expected[30].

To the model described above we add a third photoionized absorber to search for high velocity broadening as invoked in a previous study using the CCD spectra[8,12]. The $v_\sigma$ was therefore untied between all components. A highly significant improvement to the continuum-only model (Δχ$^2$/d.o.f.= 150/4, see Model 2 in Extended Data table 2) was also obtained with a highly-ionized, outflowing (at moderately relativistic velocities ~0.25$c$), and optically thick absorber (see blue line in Fig. 1). The large velocity broadening (~0.1$c$) in this latter model may explain why these features appear weak compared to the emission lines, however, an alternative reason may be variability between observations (indeed a trend in the strength of the residuals in the CCD spectra with spectral hardness was recently discovered[12]). There is some degree of degeneracy in the absorption models. The inclusion of the broadened XABS 3 component strongly decreases the significance of the other two components XABS 1-2. Longer exposures are needed to better characterize the outflow.

In principle the emission lines could also be produced by photoionized gas further away from the X-ray source. Indeed, the Fe XVII 17 Å forbidden (f) line is much stronger than the 15 Å resonance (r) line which would suggest either photoionization or resonant absorption. As SPEX does not provide a model for line-emitting gas in photoionization equilibrium, we used the *photemis* model in XSPEC[31] to create a grid of photoionization emission models with log ξ from 1.0 to 4.0 with a 0.25 step size. The best fit still supports Ne/Fe ≥ 1 abundance ratio. It is difficult to distinguish between photoionization and collisional ionization models as the results are comparable; deeper observations are necessary. Alternatively, the emission features may originate from recombination of highly ionized gas within the wind or in a distant region, which is expected if photoionization occurs. However, a recombination model does not describe the lines satisfactorily. A shock between the outflow and the low density material in the surrounding nebula is also ruled out due to the substantial X-ray brightness and the size of the X-ray source (see Extended Data Fig. 2) which is far more spatially compact (<116 pc) than the surrounding nebula (240-800 pc)[21].

**RGS-only spectral modelling**

We performed RGS-only spectral fits to check the line detection. We removed the EPIC-PN data and froze the continuum parameters to the values obtained with the simultaneous EPIC-RGS fit. We used Model 1 (with two narrow absorbers) and confirmed the need for both rest-frame emission and absorption lines. Despite the larger countrate, EPIC-PN does not enormously change the detection significance due to its poorer spectral resolution with respect to RGS in the soft X-ray band.

The individual RGS spectra for each observation of NGC 1313 X-1 show changes in absorption line strengths (see Extended Data Fig. 5). In order to study the variability of the features, we have fitted the RGS spectra for the individual exposures with a simple model consisting of one CIE line-emitting component and one XABS absorber. Absorption is detected in the first two observations with consistent parameters ($N_H$ = 2.0 ± 0.4 × $10^{22}$ cm$^{-2}$, $\xi$ = 200 ± 70 erg cm s$^{-1}$, v = -57,000 ± 500 km s$^{-1}$ ~ 0.2$c$, with significance > 3σ in total for each observation), while the emission lines are stronger in observation 3 where their flux is twice than that seen in observation 1, but the temperature is consistent at kT = 1.10 ± 0.15 keV (~1.3 × $10^7$ K). Emission lines are weaker in observation 2 and show a decrease in the ionization parameter where the Fe XVII and O VII lines (from cooler gas) are stronger than the Ne X and O VIII lines. We do not detect significant absorption in observation 3.

**RGS Line significance**

We have also confirmed the detection of each emission/absorption line by fitting the RGS spectra adopting the EPIC-RGS continuum and including a Gaussian spanning the 7–27 Å wavelength range in increments of 0.05 Å. We assumed a grid of linewidths from 500 km s$^{-1}$ (~ RGS resolution) to 75,000 km s$^{-1}$ (0.25c). In Fig. 3 we show the significance obtained adopting 500 km s$^{-1}$ and 10,000 km s$^{-1}$ line widths, confirming the lines detected with the CIE emission model. Line broadening does not have a major effect on the detection. The absorption lines have a lower significance because velocity shift is an additional parameter. The strongest feature at 11.5Å (identified as Ne IX blueshifted absorption) has a probability of 6 × $10^{-7}$ (i.e. 5σ), whether we consider it as a sum of 2 strong narrow lines or a single broad line. However, if we take into account all the trials due to the spectral resolution bins and widths, we obtain a probability ≈ 3 × $10^{-6}$, which is above 4σ. If we also include the other strong blueshifted lines that are found at exactly the same velocity, e.g. O VIII at 16.0Å and O VII at 18.0Å, then we obtain a total significance above 5σ.

In order to further check the robustness of our results, we adopt different Ne, Fe, and O abundances for the neutral absorbing gas. The neutral gas of NGC 1313 provides the bulk of the $N_H$ and may have non-solar abundances; this in principle could affect the detection of features in the soft X-ray spectra[8,32]. We have therefore re-fitted the RGS and, afterwards, the EPIC-RGS spectra, *simultaneously*, with interstellar abundances ranging from 0.1 to 2.0 × solar. No significant difference was found and the detection level of the lines is unchanged; this was expected for several reasons: the strongest features imprinted by neutral gas are expected between 22.7-23.5 Å (oxygen K edge and 1s-2p line)[33] and the lines in the ULX spectra avoid the edges. In addition, as we anticipated, the lines are narrow and their detection is not affected by the continuum-like hydrogen absorption.

We have stacked the first order RGS 1-2 spectra from the individual exposures of the same source for plotting purposes only (the stacked spectrum has much higher S/N ratio and simplifies the recognition of the lines). We have used the following advanced method to combine fluxed spectra[34]. We first created individual fluxed spectra using the SAS task *rgsfluxer* and then averaged them with the SPEX tool *rgs_fluxcombine (option 1)* for RGS 1 and RGS 2, separately. We then ran again the *rgs_fluxcombine (option 2)* to combine the stacked RGS 1 and 2 fluxed spectra into a final RGS spectrum for each ULX. Finally, we used the SPEX task *rgs_fmat* to produce the response matrix for the stacked fluxed spectrum (see the SPEX manual[13]). The stacked RGS spectra of the three ULXs are shown in Fig. 1-4, and in Extended Data Fig. 4.

**Constraints on the location and the energetics of the wind, and the black hole mass**

Here we try to place some constraints on the location of the wind seen in NGC 1313 X-1 as well as the black hole mass, using as template the parameters estimated for the extreme absorber, XABS 3 (see Extended Data table 2). The ionization parameter is defined as $\xi = L_{ion} / ( n_H R^2 ) = L_{ion} / ( n_H \Delta R\ R) \cdot (\Delta R / R)$, where $L_{ion}$ is the 1-1000 Ryd ionizing luminosity of the source, $\Delta R$, R, and $n_H$ are the thickness, the size, and the number density of the absorbing region. This leads to $R = L_{ion} / ( N_H \xi) \cdot (\Delta R / R)$, where $N_H$ is the column density. Since $\Delta R < R$, then $R < L_{ion} / ( N_H \xi) = 3 \times 10^{11}$cm, but R must also be larger than the Schwarzschild radius $R_S = 2GM / c^2$. Assuming the escape velocity to be equal to the wind speed ($0.2c$), or in other words that the wind comes from a region where its speed equals the escape velocity, we obtain $R / R_S \leq 25$, which provides an upper limit on the $M_{BH}$ of 40000 $M_{sun}$ (for a region with thickness comparable to its size). A black hole with a stellar mass, i.e. up to 100 $M_{sun}$, would imply a very thin region ($\Delta R << R$, see Extended Data Fig. 7). Throughout this calculation we adopted unity covering fraction.

It is interesting to compare the wind power to the source luminosity[35]. The outflow rate can be written as $\dot{M} = 4 \pi R^2 \rho\ v\ \Omega$, which gives a wind power $P_w = \frac{1}{2} \dot{M} v^2 = 2 \pi R^2 m_p n_H v^3 \Omega$, where $m_p$ is the proton mass and $\Omega$ the solid angle. Since $\xi = L / n_H R^2$, we get $P_w = 2 \pi L m_p v^3 \Omega / \xi$, which for component XABS 3, provides $P_w / L \approx 100\ \Omega$. This would imply a highly super-Eddington accretion rate, but could be regarded as an upper limit because a smaller outflow rate and kinetic power are obtained if either the covering fraction is lower than unity or the duty cycle is shorter[12]. On the other hand, the wind speed that we measure is a lower limit because it is only maximal for sightlines into the direction of outflow. With the present data characterized by only a few unevenly-sampled observations, we cannot accurately measure these parameters.

**NGC 5408 X-1 spectral modeling**

An X-ray image of the ultraluminous X-ray source in NGC 5408 is shown in Extended Data Fig. 3 along with another bright X-ray source (X-2, hereafter) which is covered by the RGS slit. In order to accurately estimate the RGS spectral continuum, we need to estimate the contribution from each source. We have therefore extracted the EPIC-PN spectra in two circular regions of one arcmin centered on the two sources. The background was chosen from a source-free circular region on the same chip and away from the read-out direction. The EPIC spectrum of NGC 5408 X-1 was modeled with a soft blackbody (kT = 0.14 keV ~ $1.6 \times 10^6$ K) and a power-law ($\Gamma$ ~ 2.6). The X-2 EPIC spectrum is very well modeled by a single power-law component with $\Gamma = 1.99 \pm 0.01$ and neutral column density of $5.6 \pm 0.2 \times 10^{20}$ cm$^{-2}$ consistent with the H I maps[27]. No features or residuals are detected in the X-2 EPIC spectrum; this is most likely a background AGN.

We have built up a spectral model comprising the continuum from both X-1 and X-2 EPIC spectra and applied it to the RGS stacked spectrum of NGC 5408 X-1. We searched for residual emission and absorption features as we did for NGC 1313 X-1 (i.e. using a Gaussian line stepping in wavelength). In Extended Data Fig. 6 we show the line significance obtained with 500 km s$^{-1}$ and 10,000 km s$^{-1}$ widths; the results do not strongly depend on the linewidth. The rest-frame wavelengths of some relevant transitions are labelled. Ne X and O VIII emission lines are detected each at $4\sigma$. Blueshifted absorption is also clearly detected with the strongest features detected at $3\sigma$ each (taking into account the number of velocity bins), which provides a detection $> 4\sigma$ in total for blueshifted absorption at a velocity shift of about 66,000 km s$^{-1}$.

We proceed to test physical models for the features, firstly with an isothermal emission model of gas in collisional ionization equilibrium (kT = $3.0 \pm 0.5$ keV, $\Delta\chi^2 = 129$, *d.o.f.* = 3 relative to the continuum-only fits). This is able to reproduce the Ne X and O VIII lines and the residual Fe XVII emission, but the O VII emission is underestimated (see red line in Fig. 4). To model the absorption features, we applied a photoionized absorber to the continuum components (blackbody and power-law) of NGC 5408 X-1, leaving the continuum components of X-2 unabsorbed. Most features can be reproduced with a relativistically-outflowing (v = 0.22 ± 0.01 c, $\Delta\chi^2$ = 35, d.o.f. = 3) photoionized gas model. A better, description of the spectrum, is obtained adding a cooler (kT = 0.10 ± 0.05 keV, $\Delta\chi^2$ = 15, d.o.f. = 2) CIE to fit the O VII lines and a slower (v = 0.10 ± 0.01 c, $\Delta\chi^2$ = 20, d.o.f. = 3) photoionized absorber (again only applied to the X-1 continuum components, see blue line in Fig. 4). The absorption lines have a width of 500 ± 300 km s$^{-1}$, while the emission lines are broader with $\sigma_v$ = 2000 ± 500 km s$^{-1}$, which are similar to the resolved RGS lines in NGC 1313 X-1. Solar abundances were adopted for all emission and absorption components. The highly significant, v = 0.22c, absorber in NGC 5408 X-1 has an ionization parameter $\xi$ = 50 ± 30 erg cm s$^{-1}$ and column density $N_H$ = (3.0 ± 0.4) × 10$^{20}$ cm$^{-2}$, which are lower than in NGC 1313 X-1, while their outflow velocities (v ~ 0.2c) are comparable.

**References**

[22] Strüder, L.; Briel, U.; Dennerl, K. et al. The European Photon Imaging Camera on XMM-Newton: The pn-CCD camera, *Astron. Astrophys.*, **365**, p. L18-L26 (2001)

[23] Turner, M. J. L.; Abbey, A.; Arnaud, M. et al. The European Photon Imaging Camera on XMM-Newton: The MOS cameras, *Astron. Astrophys.*, **365**, p. L27-L35 (2001)

[24] den Herder, J. W.; Brinkman, A. C.; Kahn, S. M. et al. The Reflection Grating Spectrometer on board XMM-Newton, *Astron. Astrophys.*, **365**, p. L7-L17 (2001)

[25] http://xmm.esac.esa.int/sas/

[26] Roberts, T. P.; Kilgard, R. E.; Warwick, R. S. et al. Chandra monitoring observations of the ultraluminous X-ray source NGC 5204 X-1, *Mon. Not. R. Astron. Soc.*, **371**, p. 1877-1890 (2006)

[27] Kalberla, P. M. W.; Burton, W. B.; Hartmann, Dap et al. The Leiden/Argentine/Bonn (LAB) Survey of Galactic HI. Final data release of the combined LDS and IAR surveys with improved stray-radiation corrections, *Astron. Astrophys.*, **440**, p. 775-782 (2005).

[28] Lodders, K. S. & Palme, H. Solar System Elemental Abundances in 2009, *M&PSA*, **72**, p. 5154 (2009).

[29] Nomoto, K.; Tominaga, N.; Umeda, H. et al. Nucleosynthesis yields of core-collapse supernovae and hypernovae, and galactic chemical evolution, *Nucl. Phys.*, **777**, p. 424-458 (2006)

[30] Swartz, D. A.; Ghosh, K. K.; Kajal, K. et al. The Ultraluminous X-Ray Source Population from the Chandra Archive of Galaxies, *Astrophys. J. Supp.*, **154**, p. 519-539 (2004)

[31] http://heasarc.nasa.gov/docs/software/xspec/

[32] Goad, M. R.; Roberts, T. P.; Reeves, J. N. and Uttley P. A deep XMM-Newton observation of the ultraluminous X-ray source Holmberg II X-1: the case against a 1000-M$_\odot$ black hole, *Mon. Not. R. Astron. Soc.*, **365**, p. 191-198 (2006)

[33] Pinto, C.; Kaastra, J. S.; Costantini, E.; de Vries, C. Interstellar medium composition through X-ray spectroscopy of low-mass X-ray binaries *Astron. Astrophys.*, **551**, p. 25-35 (2013).

[34] Kaastra, J. S.; de Vries, C. P.; Steenbrugge, K. C.; et al. Multiwavelength campaign on Mrk

509. II. Analysis of high-quality Reflection Grating Spectrometer spectra, *Astron. Astrophys.*, **534**, p. 37-52 (2011).

[35] King, A. L.; Miller, J. M.; Raymond, J. et al. Regulation of black hole winds and jets across the mass scale, *Astrophys. J.*, **762**, p. 103-120 (2013)

**Extended Data Table 1: Summary of the *XMM*-Newton observations.**
Exposure times after data reduction, and average de-absorbed luminosities.

**Extended Data Table 2: *XMM*-Newton EPIC-RGS spectral modelling.**
The CIE normalizations (0.3-10 keV luminosity) and temperatures (kT) are in units of $10^{38}$ erg s$^{-1}$ and keV, the XABS column densities ($N_H$) and ionization parameters (log $\xi$) are in $10^{24}$ cm$^{-2}$ and erg cm s$^{-1}$. All velocities are in km s$^{-1}$. The abundances are relative to iron, whose abundance is fixed to be solar[28]. Errorbars are given at the 1σ level.

**Extended Data Figure 1: EPIC MOS+PN stacked image of NGC 1313.**
The circular source extraction regions have a diameter of 1 arcmin. The small region at the south of X-1 is a star forming region near the galactic centre, orders of magnitude fainter across the 0.3-10keV bandpass than the ULX-1. The yellow strip is the RGS extraction region.

**Extended Data Figure 2: ACIS image of NGC 1313 X-1 and the nearby star-forming region.**
The ultraluminous X-ray source is the brightest object. The small circles have 6 arcsec radii.

**Extended Data Figure 3: EPIC MOS+PN stacked images of NGC 5408 and NGC 6946.**
The ultraluminous X-ray sources are the brightest objects in both images. Additional, nearby X-ray bright sources – mostly high-mass X-ray binaries and background active galactic nuclei – can be seen. The white circular source extraction regions have a diameter of 1 arcmin.

**Extended Data Figure 4: XMM-Newton/RGS stacked spectra of the brightest ULXs (X-1) in NGC 1313, NGC 5408, and NGC 6946.**
The rest-frame wavelengths of relevant transitions are labelled. The spectra have been re-binned for display purposes. Errorbars are given at the 1σ level.

**Extended Data Figure 5: XMM-Newton RGS spectra and best-fitting model to each observation of NGC 1313 X-1.**
The rest-frame wavelengths of the most relevant emission lines (green) and the blueshifted absorption lines (blue) are labelled. Obs 1 shows both absorption and emission lines. Obs 2 is dominated by absorption, while obs 3 shows mostly emission features. Errorbars on data are given at the 1σ level.

**Extended Data Figure 6: Significance of the features in the NGC 5408 X-1 RGS stacked spectrum.**
Negative values refer to absorption lines (see also Fig. 3 for NGC 1313 X-1). The solid and dashed curves show the line significance obtained with 500 km s$^{-1}$ and 10,000 km s$^{-1}$ widths, respectively.

**Extended Data Figure 7: Constraints on the location of the extreme absorber, XABS 3.**
The white area refers to the acceptable values between the Schwarzschild radius (bottom oblique line), the $\xi = L / n_H R^2$ equation (dotted horizontal line), and the radius assuming the escape velocity to be equal to the the wind speed (0.2c, top oblique line).

| Source | Observation ID | $t_{TOT}$ (ks) | $L_{0.3-10\,keV}$ (erg s$^{-1}$) |
|---|---|---|---|
| NGC 1313 X-1 | 0405090101, 0693850501, 0693851201 | 345.6 | $1.04 \times 10^{40}$ |
| NGC 5408 X-1 | 0302900101, 0500750101, 0653380201, 0653380301, 0653380401, 0653380501 | 644.9 | $2.01 \times 10^{40}$ |
| NGC 6946 X-1 | 0691570101 | 110.0 | $0.97 \times 10^{40}$ |

| **Model 1 with narrow lines** | | | |
|---|---|---|---|
| Parameter | CIE | XABS 1 | XABS 2 |
| $N_{CIE}$ or $N_{H,XABS}$ | $3.1 \pm 0.4$ | $1.5 \pm 0.3 \times 10^{-2}$ | $5 \pm 1 \times 10^{-3}$ |
| $T_{CIE}$ or $\xi_{XABS}$ | $0.85 \pm 0.03$ | $2.20 \pm 0.04$ | 2.20 coupled |
| $v_\sigma$ | 10 ($<500$) | 10 coupled | 10 coupled |
| $v_{outflow}$ | $\equiv 0$ | 0 ($> -3 \times 10^3$) | $-6.5 \pm 0.2 \times 10^4$ |
| O/Fe | $1.0 \pm 0.2$ | 1.0 coupled | 1.0 coupled |
| Ne/Fe | $1.8 \pm 0.4$ | 1.8 coupled | 1.8 coupled |
| Mg/Fe | $2.1 \pm 0.5$ | 2.1 coupled | 2.1 coupled |
| $\Delta\chi^2$, $\Delta C_{stat}$, d.o.f. | 87, 114, 6 | 130, 65, 3 | 48, 20, 3 |

| **Model 2 with broad lines** | | | | |
|---|---|---|---|---|
| Parameter | CIE | XABS 1 | XABS 2 | XABS 3 |
| $N_{CIE}$ or $N_{H,XABS}$ | $3.0 \pm 0.4$ | $3.8 \pm 1.3 \times 10^{-3}$ | $3.1 \pm 1.0 \times 10^{-3}$ | $1.1 \pm 0.2$ |
| $T_{CIE}$ or $\xi_{XABS}$ | $0.80 \pm 0.03$ | $2.29 \pm 0.09$ | 2.29 coupled | $4.55 \pm 0.22$ |
| $v_\sigma$ | $1250 \pm 600$ | 10 ($<20$) | 10 ($<20$) | $3.0 \pm 1.5 \times 10^4$ |
| $v_{outflow}$ | $\equiv 0$ | 0 ($> -3 \times 10^3$) | $-3.9 \pm 0.2 \times 10^4$ | $-7.5 \pm 1.5 \times 10^4$ |
| O/Fe | $1.38 \pm 0.16$ | 1.38 coupled | 1.38 coupled | 1.38 coupled |
| Ne/Fe | $3.88 \pm 0.91$ | 3.87 coupled | 3.87 coupled | 3.87 coupled |
| Mg/Fe | $3.33 \pm 0.50$ | 3.33 coupled | 3.33 coupled | 3.33 coupled |
| $\Delta\chi^2$, $\Delta C_{stat}$, d.o.f. | 86, 107, 6 | 30, 15, 4 | 12, 10, 4 | 150, 41, 4 |

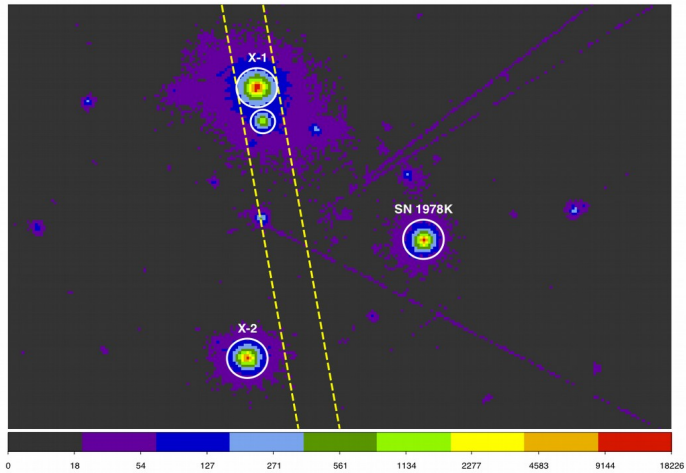

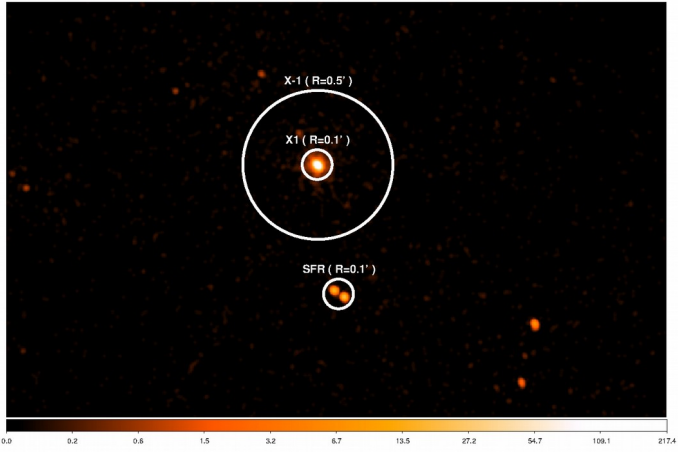

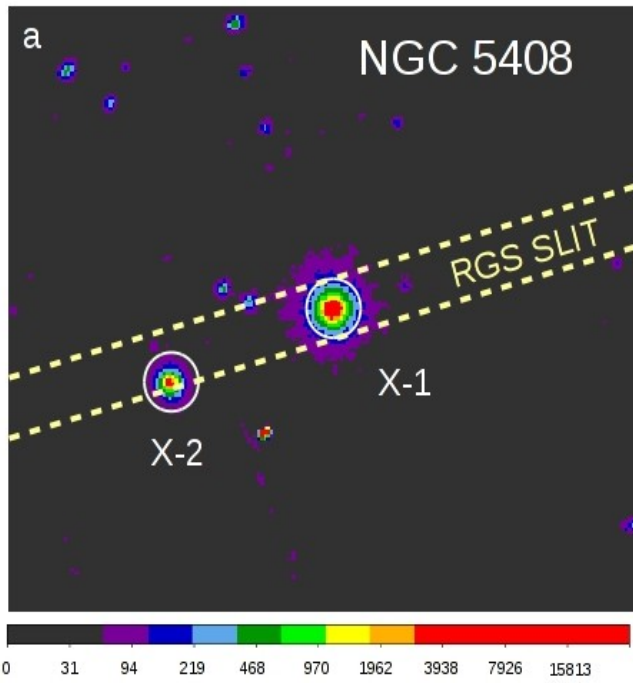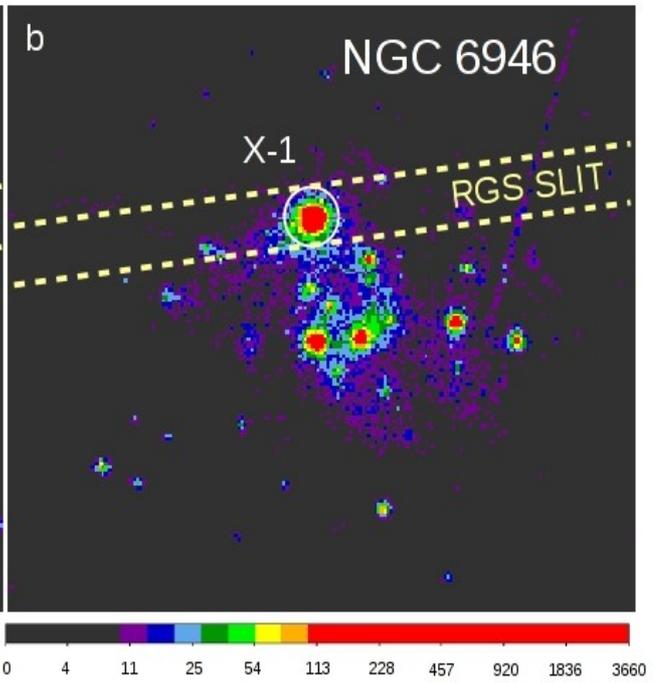

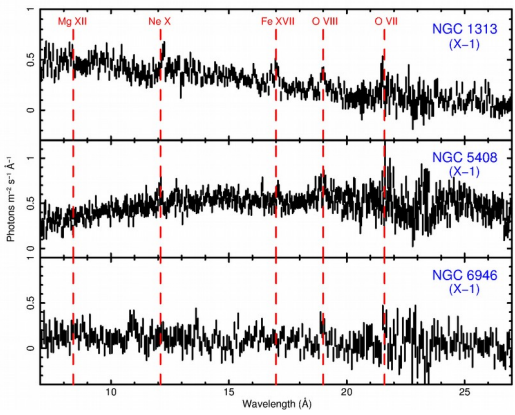

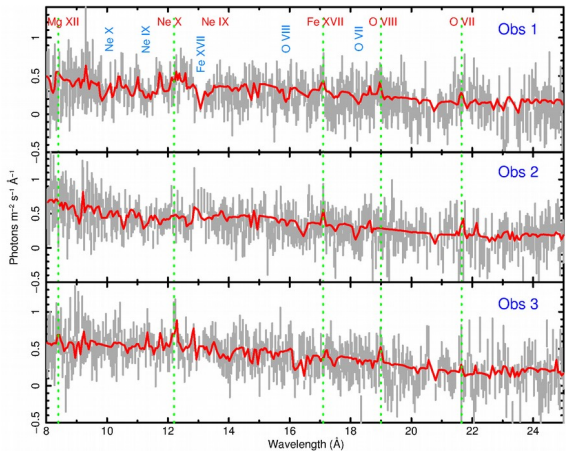

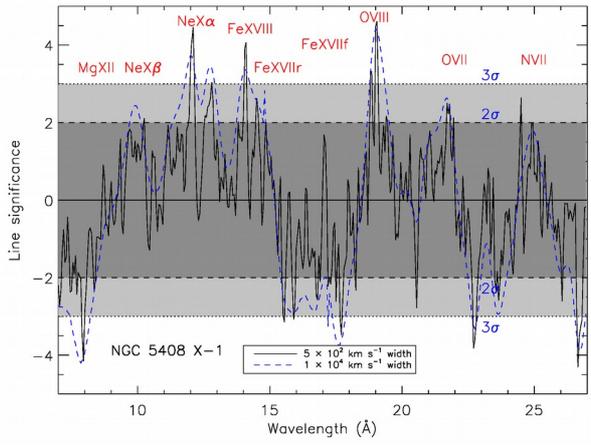

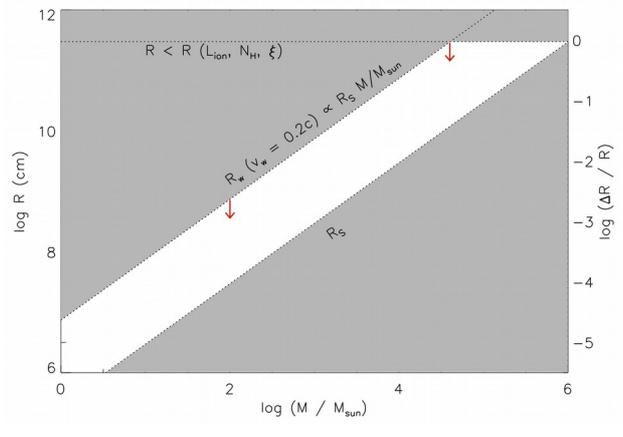